\documentclass[a4paper,fleqn,usenatbib]{mnras}

\usepackage[T1]{fontenc}
\usepackage{ae,aecompl}

\usepackage{graphicx}	
\usepackage{amsmath}	
\usepackage{amssymb}	
\usepackage{threeparttable}
\usepackage{tablefootnote}
\title[BLR dynamics of NGC 3227]{The dynamics of the broad-line region in NGC 3227}

\author[Devereux]{
Nick Devereux$^{1}$\thanks{E-mail: devereux@erau.edu}
\\
$^{1}$Embry-Riddle Aeronautical University, 3700 Willow Creek Road, Prescott, AZ 8301, USA\\
}

\date{Accepted XXX. Received YYY; in original form ZZZ}

\pubyear{2020}

\begin{document}
\label{firstpage}
\pagerange{\pageref{firstpage}--\pageref{lastpage}}
\maketitle

\begin{abstract}

Archival {\it Hubble Space Telescope} ({\it HST}) observations of the Seyfert 1 nucleus of NGC 3227 obtained with the Space Telescope Imaging Spectrograph (STIS) are re-examined
in order to constrain a viable photoionization model for the broad-line region (BLR). The results imply that the BLR is a partially ionized, dust-free, spherical shell that is collapsing,
supersonically, at the free-fall velocity due to its proximity to a supermassive black hole. The  BLR is ionization bounded at the outer radius, coincident with the dust reverberation radius, and transforms into an X-ray emitting plasma inside the Balmer reverberation radius as the central UV--X-ray source is approached. Only 40 M$_{\sun}$ of Hydrogen are 
required to explain the Balmer emission line luminosity, but it is compressed by gravity into a column measuring 5.5 ${\times}$10$^{24}$ atoms cm$^{-2}$.
Assuming radiatively inefficient accretion, the X-ray luminosity requires ${\sim}$ 10$^{-2}$ M$_{\sun}$ yr$^{-1}$. 
However, the mass inflow rate required to explain the luminosity of the broad H${\alpha}$ emission line is ${\sim}$ 1 M$_{\sun}$ yr$^{-1}$. The very large disparity between these two estimates indicates that 99\% of the inflowing gas must be re-directed into an outflow, and on a very short timescale corresponding to ${\sim}$ 40 years. Alternatively,
the radiative efficiency of the inflow has been overestimated, or the X-ray luminosity has been underestimated; a distinct possibility if the BLR is indeed Compton thick.
 
\end{abstract}

\begin{keywords}
galaxies: Seyfert, galaxies: individual (NGC3227)
\end{keywords}


\section{Introduction}

Fortuitously, nature has provided a relatively unobscured view of the Seyfert 1 nucleus in the spiral galaxy NGC 3227, at least at visible wavelengths \citep{Crenshaw_2001}. This circumstance is afforded by the favourable inclination of both the host galaxy and the unresolved dust torus \citep{Alonso_Herrero_2011}. Consequently, observations obtained with the Space Telescope Imaging Spectrograph (STIS) aboard the {\it Hubble Space Telescope} ({\it HST}) reveal the emergent visible spectrum of the nucleus with unprecedented clarity and at an angular resolution that effectively excludes the galaxy starlight with an efficacy that is presently unmatched by any other telescope on or off the planet \citep[][hereafter D13]{Devereux_2013}.

According to \cite{Rubin_1968}, who published the first visible spectrum, NGC 3227 was among the original group of objects identified by Seyfert, but it was not included in \cite{Seyfert_1943}.  
Subsequently, it became apparent that the broad Balmer emission lines are time-variable \citep[][and references therein]{Pronik2009, Rosenblatt_1992, Rosenblatt_1994}. 
Modern measurements of the time-delay between correlated changes in the brightness of the broad H${\beta}$ emission line and the adjacent continuum led to a measure of 
physical {\it size} corresponding to ${\sim}$ 1 light-day \citep{De_Rosa_2018}. However,  because only ${\sim}$ 10\% of the broad H${\beta}$ line flux actually varies, it is not clear what 
this {\it size} refers to exactly. Besides, there are other physical properties that remain to be established such as gas density, volume filling factor, 
metallicity, and geometry, parameters that are central to understanding the current dynamical state of the BLR, its origin and subsequent evolution.

Broad emission lines are a defining characteristic of active galactic nuclei (AGN), and a considerable literature has evolved over the past several decades identifying them, in particular, 
with dense broad-line clouds. Intriguingly, dense clouds do appear to be present in the BLR of NGC 3227, but they are not as dense as previously imagined, and they 
are few in number, at least along our line of sight to the central X--ray source \citep{Lamer_2003,Beuchert_2015,Turner_2018}.  Consequently, an alternative 
explanation is explored here whereby the broad Balmer emission lines are identified with an H$^+$ region that is photoionized by the central 
UV--X-ray source, akin to the ${\ion{H}{ii}}$ regions commonly observed in association with planetary nebulae, supernova remnants and massive star forming regions, 
but with a much higher ionization parameter.

{\it HST} observations reveal a distinctly triangular shape for the broad H${\alpha}$ emission line (D13) that has apparently persisted for three decades
since the first published spectrum \citep{Rubin_1968}.
The shape represents the convolution of the Balmer emissivity with a velocity field and a geometry for the ionized gas, none of which are presently known. 
This is because the BLR is unresolved (< 17 ${\times}$ 10$^{-3}$ arc sec ) and is 
likely to remain so for many decades to come. However, some useful constraints can be deduced from existing observations.
Firstly, most, if not all, of the H ionizing photons produced by the central UV-X--ray source are accounted for by the broad Balmer emission line luminosity. 
Therefore, the H$^+$ region producing the broad Balmer emission lines seen in NGC 3227 has a very high covering factor, suggesting an approximately 
spherical distribution for the ionized gas (D13). Secondly, the inner radius of the H$^+$ region emitting the Balmer emission lines coincides with the Balmer reverberation radius 
${\sim}$ 1 light-day \citep{De_Rosa_2018}, whereas the outer radius of the H$^+$ region is coincident with the dust reverberation radius corresponding to ${\sim}$ 20 light-day 
\citep{Suganuma_2006, Koshida_2014}. Thus, the BLR in NGC 3227 is essentially a photoionized dust-free spherical shell of H$^+$ gas with filling factor ${\sim}$ 1 (D13).
Insofar as the outer boundary of the H$^+$ region in NGC 3227 is coincident with the dust reverberation radius, the BLR model is similar to that proposed by \cite{Netzer_1993}, but differs with respect to the notion of clouds. 

Estimates for the central black hole (BH) mass based on reverberation mapping have been steadily declining in recent years to a value M$_{\bullet}$ = 4 ${^{+2}_{-2}}$ ${\times}$ 10$^{6}$ M${_{\sun}}$ \citep[][and references therein]{De_Rosa_2018,Peterson_2004}, similar to that obtained using the prescription of \cite{Greene_2005} but smaller than prior estimates based on stellar and gas kinematics \citep{Davies_2006,Hicks_2008}. Nevertheless, combining the most recent BH mass estimate together with the bolometric luminosity, L$_{bol}$ = 1.9 ${\times}$ 10$^{43}$ erg/s  \citep{Winter_2012}, corrected for the slightly larger distance of 20.8 Mpc adopted here \citep{Tully1988}, leads to an Eddington luminosity ratio L$_{bol}$/L$_{Edd}$ ${\sim}$ 4 
${\times}$ 10$^{-2}$. The diminutive value for this ratio implies that any matter divested of angular momentum will inevitably fall inwards, supersonically, at the free-fall velocity, because the radiation force is too small, by over one order of magnitude, to withstand the gravitational force of the central BH. Observationally, the luminosity of the central UV--X-ray source implies a ballistic mass inflow rate ${\ge}$ 10$^{-2}$ M${_{\sun}}$/yr assuming radiatively inefficient accretion (D13).  

High angular resolution (0.2 arc sec) spectroscopy of the BLR in NGC 3227 has been obtained with STIS on two separate occasions (D13). The first in 1999 used the G750M grating to obtain a well sampled (0.56 {\AA} pixel$^{-1}$) spectrum of the broad H${\alpha}$ emission line. On the second occasion in 2000, STIS was used to obtain a less well sampled (4.92 {\AA} pixel$^{-1}$) spectrum using the G750L grating, in addition to contemporaneous spectra with the G430L, G230L and G140L gratings. NGC 3227 is just one of a few AGN with {\it HST} spectra spanning the visible to the UV \citep{Spinelli_2006}.

This paper reports the results of using the computer code {\scriptsize XSTAR} \citep{Kallman_2001} to model photoionization of the H$^+$ region in an effort to constrain the physical and dynamical state of the BLR gas. Of particular interest is whether the combination of outward radiation pressure and gas pressure is sufficient to overcome the inward gravitational force given the proximity of the BLR to a several million solar mass BH.
The present analysis involves modelling the shape and luminosity of the broad H${\alpha}$ emission line with two important refinements on the previous work (D13). First, rather than assuming a Balmer emissivity, it is calculated explicitly using the photoionization code {\scriptsize XSTAR}
for a parameter space defined by a variety of densities and radial distributions for the ionized gas. 
Secondly, the analysis explores two plausible kinematic models for the BLR gas that can potentially be distinguished, namely randomly oriented circular orbits and radial motion at the escape velocity. On the one hand, inferences regarding the kinematic state of the BLR gas that are based on reverberation mapping studies
are inconclusive \citep{De_Rosa_2018}.  On the other hand, one would expect a measurable, factor of ${\sqrt{2}}$,
difference in the full-width at half-maximum (FWHM) of emission lines produced by BLR gas that is distributed in randomly oriented circular orbits versus radial motion at the escape velocity, with the former narrower than the latter, all other parameters being equal. An objective of the present work, therefore, is to explore to what extent the distinctive shape observed for the broad H${\alpha}$ emission line seen in NGC 3227 can distinguish between different kinematic models for the BLR gas.

The outline for the paper is as follows. In the next section, a model for the emergent spectrum of the H$^+$ region, synonymous now with the BLR, is computed using
the photoionization code {\scriptsize XSTAR}. The emergent spectrum depends in large part on the shape and amplitude of the unobservable H ionizing continuum, models for which can be constrained using prior {\it HST} observations (D13), as explained in Section 2.1. The dust extinction to the intrinsic continuum can be inferred by comparing it with the emission-line-free continuum observed with {\it HST}, as explained in Section 2.2. It is demonstrated in Section 2.3 that the H${\alpha}$/H${\beta}$ emission line ratio observed with {\it HST} is consistent with 
little or no visible dust extinction to, or within, the BLR. Consequently, the shape and luminosity of the bright H${\alpha}$ emission line can be used to constrain {\scriptsize XSTAR} photoionization models as explained in Section 2.4. Having determined the overall geometry and ionization structure for the BLR allows the dynamics to be explored as discussed in Section 3. Conclusions follow in Section 4.

\section{Photoionization modelling}

\subsection{The intrinsic ionizing continuum}

Regrettably, the intrinsic H ionizing continuum is unobservable due to the large column of intervening neutral H gas.  However, the shape and amplitude can be constrained because it must connect sensibly with the observed visible continuum and produce enough ionizing photons to explain the broad H${\alpha}$ emission line luminosity
(see equation 1 in D13). \cite{Vasudevan_2009} proposed a H ionizing continuum for NGC 3227 consisting of a modified blackbody and a power law. But, numerically integrating that continuum produces about a factor of 6 too few H ionizing photons to explain the broad Balmer emission line luminosity, and so it was not considered further. Naively fitting a power law of the form $f_{\nu}$ ${\propto}$ ${\nu}^{-\alpha}$ 
to the emission-line-free visible continuum observed with the G750L and G430L gratings (see Fig. 1) yields ${\alpha}$ = 1.3, and a factor of ${\sim}$ 3 too few H ionizing photons.
One can therefore infer that the intrinsic ionizing continuum must have a shallower slope, similar to that produced by an advection dominated accretion flow (ADAF), a possibility anticipated by \cite{Markowitz_2009}. Computing an ADAF continuum explicitly for NGC 3227 is beyond the scope of the present paper.  Instead, the ADAF continuum computed for NGC 3998 by \cite{Nemmen_2014}\footnotemark \footnotetext{https://figshare.com/articles/Spectral\_models\_for\_low-luminosity\_active\_galactic\_nuclei\_in\_LINERs/4059945} was adopted to represent the shape of the H ionizing continuum in NGC 3227. The spectrum was subsequently scaled by a factor of 6.31 to produce an ionizing luminosity, $L_{ion}$ = 5.67 ${\times}$ 10$^{42}$ erg s$^{-1}$, integrated between 1 and 1000 Ryd, yielding 4.76 ${\times}$ 10$^{52}$ H ionizing photons s$^{-1}$, superseding a prior estimate (D13). Figure 1 illustrates the intrinsic H ionizing continuum in the context of the emission-line-free visible--UV continuum of NGC 3227 observed with STIS. The intrinsic and observed continua intersect at visible wavelengths but deviate systematically at shorter wavelengths indicative of significant reddening.

\begin{figure}
	\includegraphics[width=\columnwidth]{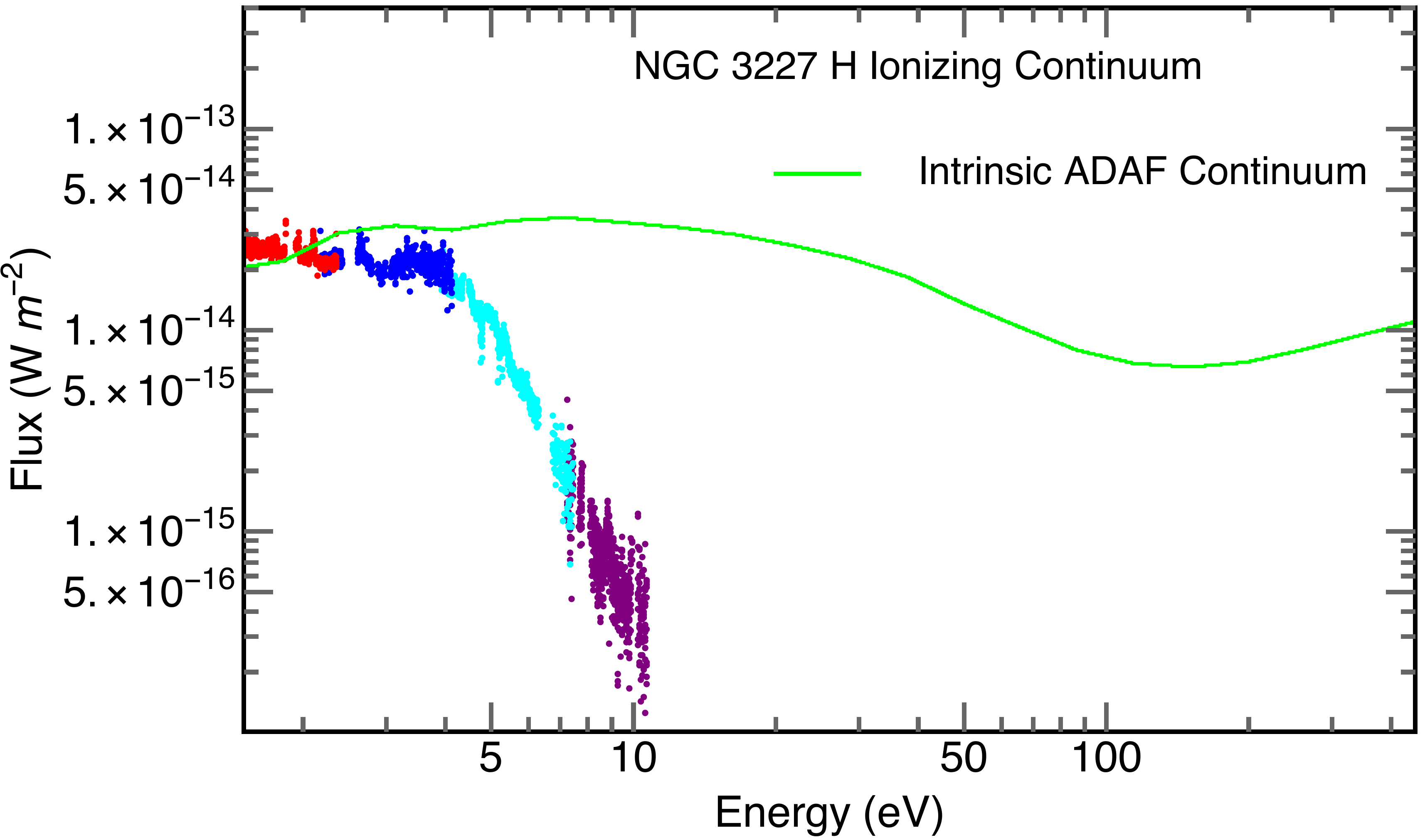}
     \caption{The visible--UV emission line-free continuum of NGC 3227. The observed continuum, defined by contemporaneous {\it HST}/STIS observations obtained with the G750L, G430L, G230L and G140L gratings, is depicted by red, blue, indigo and purple dots, respectively. The solid green line represents the intrinsic ionizing continuum (see Section 2 for details). The ordinate and the abscissa are in units of W/m$^2$ and eV, respectively.}

   \label{fig:example_figure}
\end{figure}

\subsection{UV dust extinction to the central continuum source}

The dust extinction A$_{\lambda}$, illustrated in Fig. 2 as a function of wavelength ${\lambda}$, was derived by comparing the visible--UV emission line-free continuum observed with {\it HST} to the incident ADAF continuum described in the previous section.  A$_{\lambda}$ is represented by a function defined most recently by \cite{Devereux_2019}. As Fig. 2 illustrates, the dust extinction exhibits a steep inverse dependence on wavelength in the UV.  A least squares fit to the extinction derived using just the emission-line-free continua observed with the G140L and G230L gratings yields 

\begin{equation}
A_{\lambda} = \frac{(9129 \pm 26)}{\lambda(\textrm\AA)} - (2.34 \pm 0.01) ~~~~\textrm{mag} ~~~~1150 {\leq} {\lambda(\textrm\AA)} {\leq} 3180
\end{equation}
The rest-frame dust extinction described by equation 1 is not quite as steep in the UV compared to a prior and independent determination by \cite{Crenshaw_2001}, but there is agreement on the conspicuous absence of the 2175 {\AA} feature seen in Galactic extinction curves \citep{Cardelli_1989}. Fig. 2 indicates that the observed visible continuum is dominated by starlight at the longest wavelengths and therefore provides an increasingly unreliable measure of the dust extinction to the central continuum source at ${\lambda}$ ${\geq}$ 3500 {\AA}.

\begin{figure}
	\includegraphics[width=\columnwidth]{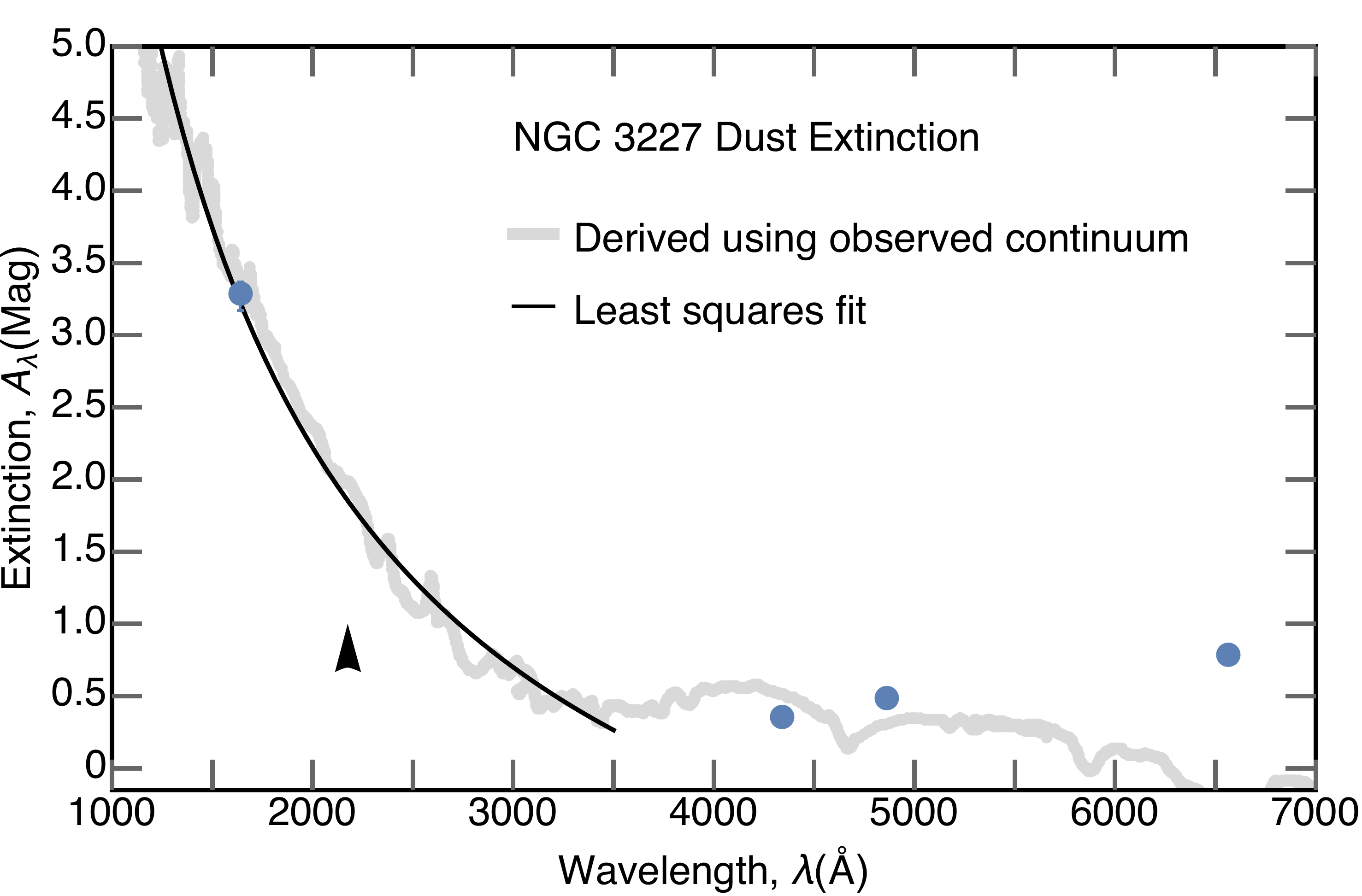}
     \caption{Derived dust extinction to the continuum source. The light-grey line identifies the rest-frame dust extinction derived by comparing the visible--UV emission-line-free continuum of NGC 3227 observed with {\it HST}/STIS to the intrinsic ADAF continuum (see Section 2.1 for details). A least squares fit to the extinction derived using just the G140L and G230L emission-line free continua is depicted by the black line. The black arrowhead identifies the wavelength expected for the 2175 {\AA} feature. Blue dots represent the dust extinction to the H${\alpha}$, H${\beta}$, H${\gamma}$ and \ion{He}{ii} ${\lambda}$1640 emission lines (see Section 2.4.2 for details). The ordinate shows the absolute extinction in units of mag versus the wavelength plotted on the abscissa in units of Angstroms.}

   \label{fig:example_figure}
\end{figure}

\subsection{Dust extinction to the broad Balmer emission lines}

Comparing contemporaneous observations of the H${\alpha}$ and H${\beta}$ emission lines, described previously in D13, potentially reveals the dust extinction to 
the H$^+$ region from which the lines originate. Furthermore, the Balmer lines are broad and sufficiently resolved to examine the dust extinction within
the H$^+$ region by adopting velocity as a proxy for radial distance from the central BH. The lower panel in Figure 3 illustrates that the observed 
mean H${\alpha}$/H${\beta}$ ratio, computed by interpolating the H${\alpha}$
flux to the same velocities at which the H${\beta}$ flux was measured, is 3.1 ${\pm}$ 0.7 and not statistically significantly different from the 
ratio of 2.7 expected for a low density 10$^{4}$ cm$^{-3}$ H$^+$ region, and is consistent, to first order, with an insignificant colour excess 
corresponding to $E(B-V)$ ${\le}$ 0.2 mag. Thus, both the shape and luminosity of the broad H${\alpha}$ emission line are, to first order, unaffected by dust extinction.
That makes them useful observational constraints with which to distinguish between various {\scriptsize XSTAR} photoionization models. 

\begin{figure}
	\includegraphics[width=\columnwidth]{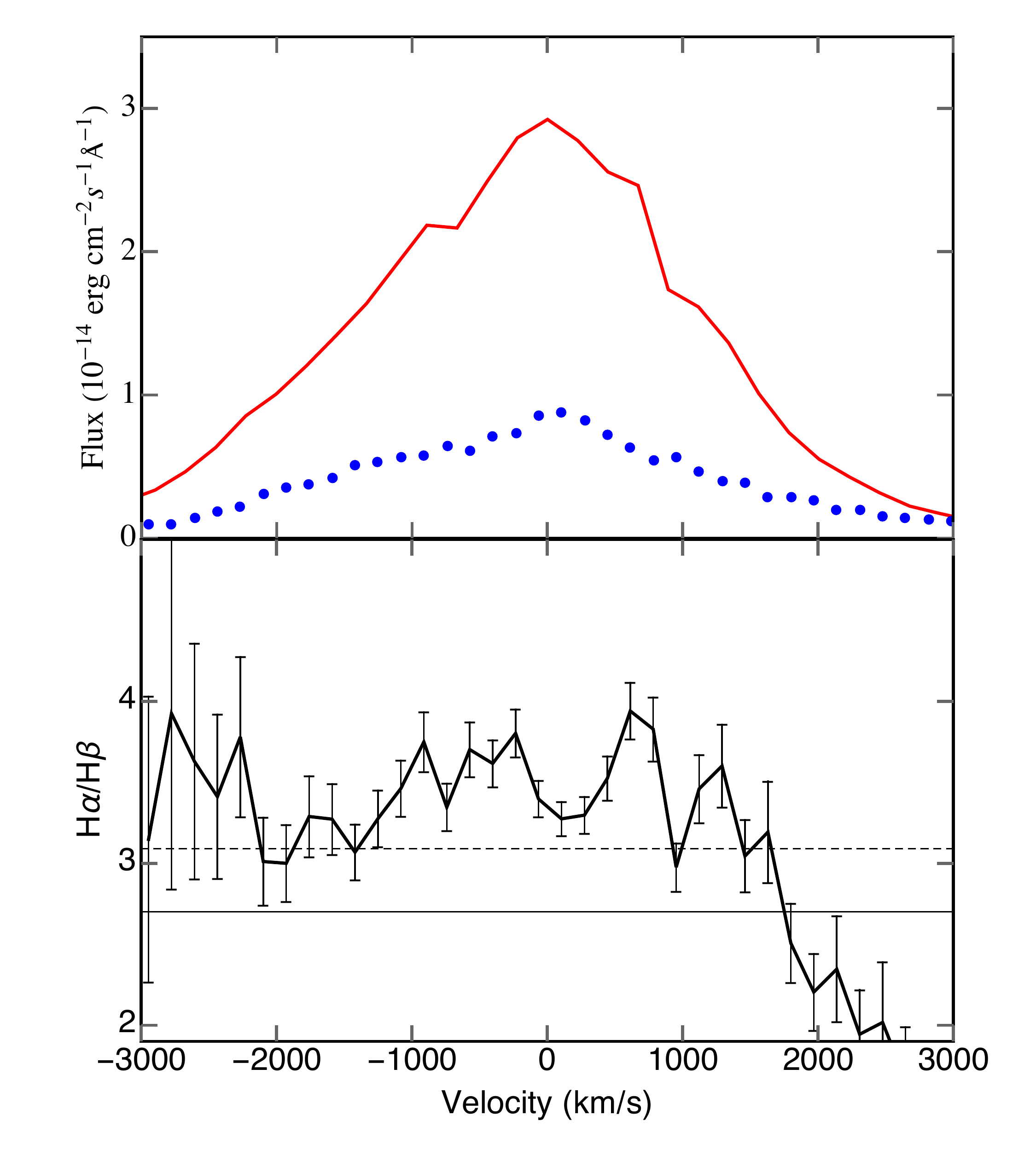}
     \caption{(Top panel) H${\alpha}$ (red-solid line) and H${\beta}$ (blue-dotted line) emission lines as observed, contemporaneously, with {\it HST}/STIS. The ordinate indicates the continuum-subtracted emission line flux density in units of erg cm$^{-2}$ s$^{-1}$ {\AA}$^{-1}$.  (Lower panel) The ordinate indicates the observed H${\alpha}$/H${\beta}$ flux density ratio having sampled the H${\alpha}$ emission line at the same velocities as measured for H${\beta}$. The solid horizontal black line represents the ratio 2.7 expected for photoionization of a low density nebula. The horizontal dashed line identifies the observed unweighted mean ratio 3.1. Error bars denote flux density measurement uncertainties only. The abscissa represents radial velocity in km s$^{-1}$ relative to the rest frame of the broad H${\alpha}$ line (see Section 2.3 for details). }

   \label{fig:example_figure}
\end{figure}

\begin{figure}
	\includegraphics[width=\columnwidth]{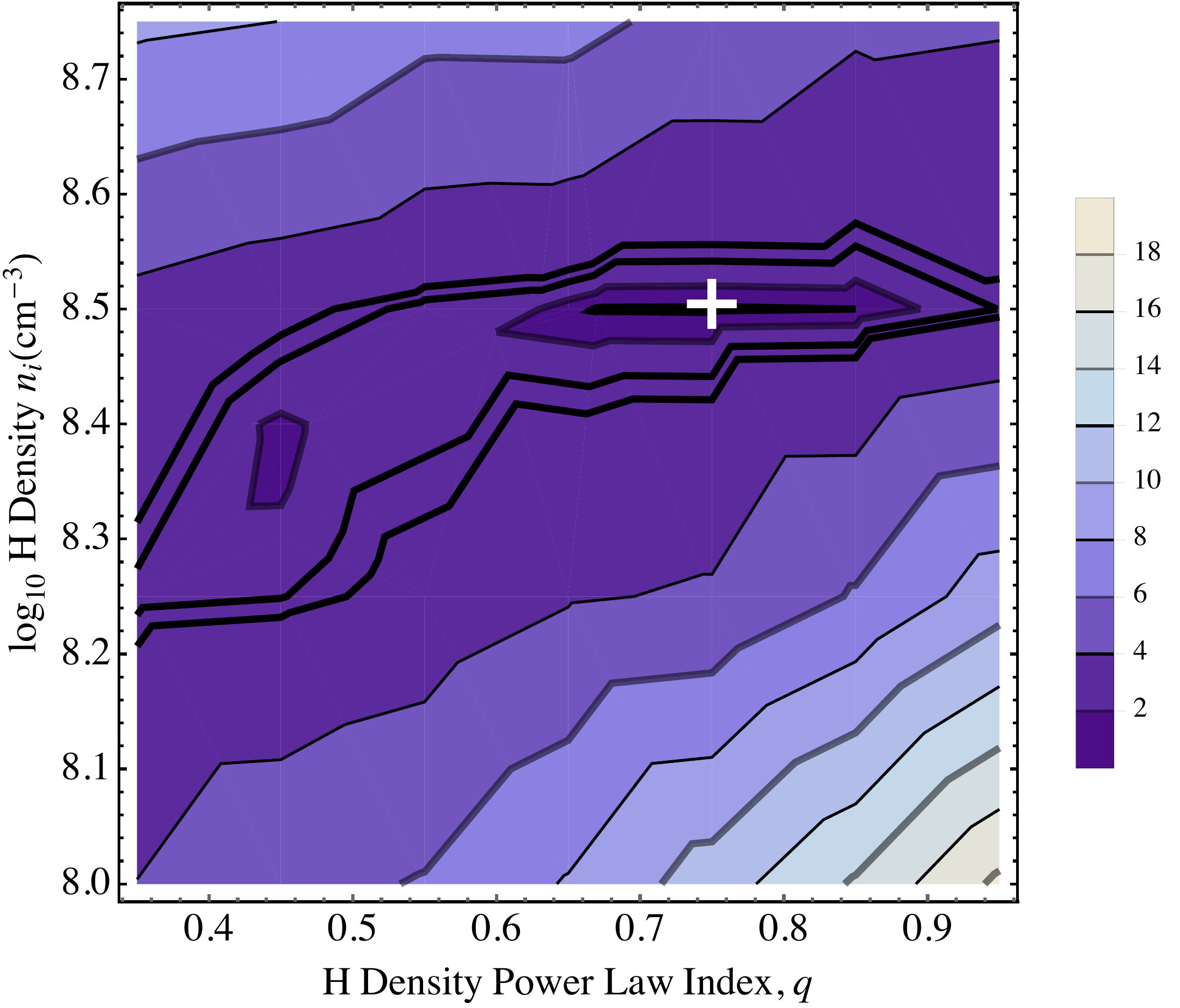}
     \caption{{\scriptsize XSTAR} photoionization model results depicted as contours representing the value of the reduced chi-squared statistic, ${\chi}_{red}^2$, computed by comparing the shape of the observed broad H${\alpha}$ emission line with model ones.  The four smallest contours represent ${\chi}_{red}^2$ =1.75, 2, 2.3, and 4 whereas the background colour corresponds to values of ${\chi}_{red}^2$ in linear steps of 2 as indicated by the colour-bar. The grid of photoionization models is defined by the power law index ${q}$, plotted on the abscissa, describing the radial distribution for a spherically symmetric ball of gas, and the gas density at the inner radius, log$_{10}$ ${n}_{i}$ (cm$^{-3}$), plotted on the ordinate, where the inner radius, $r_i$ = 10$^{-3}$ pc, is approximately coincident with the Balmer reverberation radius. The white plus sign identifies the location of ${\chi}_{red}^2$ minimum corresponding to the {\scriptsize XSTAR} model parameters log$_{10}$ ${n}_{i}$  (cm$^{-3}$) = 8.5,  log$_{10}$ ${\xi}_i$ (erg cm s$^{-1}$) = 3.3, and ${q}$ = 0.75. }
   \label{fig:example_figure}
\end{figure}

\subsection{{\scriptsize XSTAR} photoionization modelling results}

{\scriptsize XSTAR} is a computer code that predicts the emergent spectrum for a photoionized gas.\footnotemark
\footnotetext{https://heasarc.gsfc.nasa.gov/xstar/docs/html/xstarmanual.html} Version 2.39 of {\scriptsize XSTAR} was used to simulate the effect of photoionizing spherically symmetric balls of neutral H gas with various radial number density distributions described by ${n}(r)$=${n}_i(r/r_i)^{-q}$ where ${n}_i$ 
is the number density at an inner radius $r_i$, and ${r}$ is the radial distance from the photoionizing source. The inner radius is parameterized within {\scriptsize XSTAR} in terms of the ionization parameter at the inner radius, ${\xi}_i$ = $L_{ion}$/(${n}(r_{i}) r_{i}^{2}$) with $L_{ion}$ defined in Section 2.1. Physically, the inner radius represents the transition from a photoionized gas to an X-ray emitting plasma as the central UV--X-ray source is approached. Empirically, $r_i$ has been found to coincide with the Balmer reverberation radius (D13). 

Figure 4 presents a grid of models spanning 0.35 ${\le}$ $q$ ${\le}$ 0.95 and 8 ${\le}$ log$_{10}$ ${n}_{i}$  (cm$^{-3}$) ${\le}$ 8.75 at $r_{i}$ = 10$^{-3}$ pc. All of the models 
considered are ionization bounded with unity covering and filling factors. No constraint was imposed on the gas pressure, the gas temperature or the H column density.  Each photoionization model employs solar abundances limited to H, He, C, N, O, Ne, Mg, S and Fe, elements for which emission lines are seen in the {\it HST} spectra. Limiting the number of elements hastens the computations with no significant difference in results compared to those obtained using the complete suite.

Each photoionization model produces a unique radial dependence for the Balmer emissivity that serves as a probability distribution for computing a model H${\alpha}$ emission line. Briefly, the collective H${\alpha}$ emission expected from a spherical distribution of points was created by sampling the probability distribution at a variety of random radii, ${r}$, that, assuming Newtonian kinematics, translates into a number distribution of velocities which is subsequently projected into a spherical coordinate system by randomly sampling azimuth angles 0 ${\le}$ ${\phi}$ ${\le}$ 2${\pi}$ in the $x-y$ plane and elevation angles  -${\pi}$/2 ${\le}$ ${\theta}$ ${\le}$ ${\pi}$/2 measured from the $x-y$ plane to the $z$-axis. All the points were then binned into a histogram of radial velocities at the same velocity resolution as the {\it HST} observations.  

A spherically symmetric model that is randomly sampled in ${\sin\theta}$ results in a uniform surface density of points producing broad emission lines that are rectangular in shape \citep{Devereux_2007, Pancoast_2011}. However, when observed at the highest spectral resolution with {\it HST}, the broad H${\alpha}$ emission line appears to be distinctly triangular in shape which, as demonstrated previously (D13), can be replicated by randomly sampling ${\theta}$ rather than ${\sin\theta}$. This causes points to congregate at the poles of the sphere resulting in an order of magnitude higher surface density compared to the equator. In the context of radial motion, points located near the poles are moving essentially perpendicular to the line-of-sight, producing a peak in the histogram of radial velocities at ${\sim}$ 0 kms$^{-1}$. The height of the peak, equivalent to the intensity of the emission line at zero velocity, is determined by the outer radius of the spherical volume containing the points. Since the {\scriptsize XSTAR} photoionization models are ionization bounded, 
the outer radius is marked by a sharp decline in the Balmer emissivity coincident with the Str{\" o}mgren radius  \citep{Stromgren_1939, Osterbrock_1989}. 
Similarly, the inner radius is caused by a downturn in the Balmer emissivity caused by an increase in ionization as the central UV--X-ray source is approached. Thus, the inner and outer radii are not free parameters as they are in most other BLR models in the published literature. Consequently, there are just three free parameters; ${\xi}_i$, ${n}_{i}$, and $q$ to be constrained by comparing the observed broad H${\alpha}$ emission line, shape and luminosity, with model ones using the ${\chi}^2$ statistic.

\begin{figure}
	\includegraphics[width=\columnwidth]{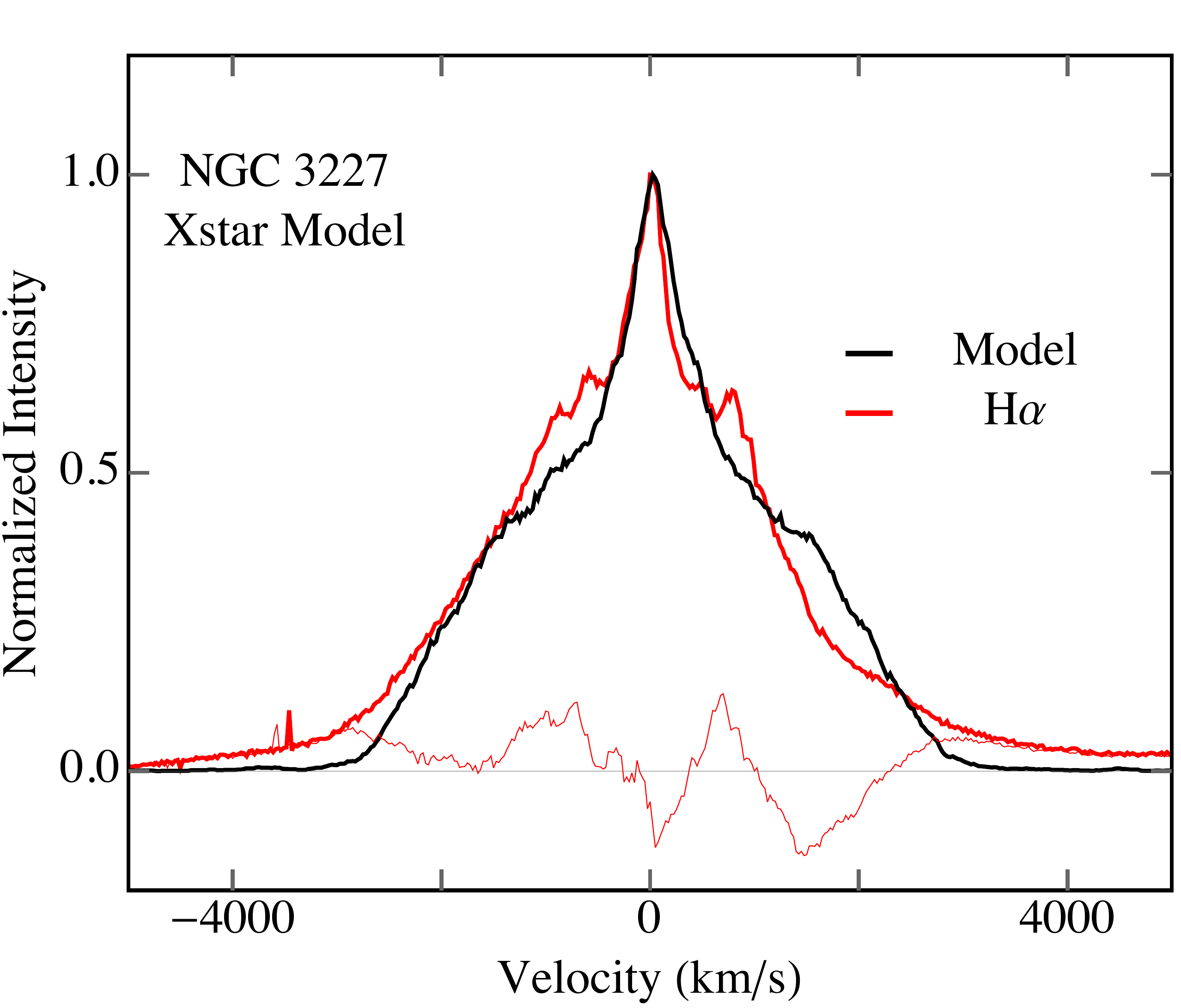}
     \caption{Comparison of the observed H${\alpha}$ emission line (red) with the model one (black) obtained at the location of minimum ${\chi}_{red}^2$ = 1.5 in Fig. 4. The 
 model H${\alpha}$ line represents the collective emission of 20458 points moving at the escape velocity in a spherical volume defined by 10$^{-3}$ ${\le}$ $r$(pc) ${\le}$ 0.02, -${\pi}$/2 ${\le}$ ${\theta}$ ${\le}$ ${\pi}$/2, and 0 ${\le}$ ${\phi}$ ${\le}$ 2${\pi}$. The probability distribution describing the number of points at each radius is identified with the Balmer emissivity of the {\scriptsize XSTAR} photoionization model with parameters log$_{10}$ ${n}_{i}$  (cm$^{-3}$) = 8.5, log$_{10}$ ${\xi}_i$ (erg cm s$^{-1}$) =  3.3, and $q$ = 0.75. The H${\alpha}$ emission line was observed with {\it HST}/STIS in 1999 using the G750M grating. See D13 for details.} 

   \label{fig:example_figure}
\end{figure}

The photoionization modelling results are summarized in Fig. 4 which displays contours of the reduced ${\chi}^2$ statistic
\begin{equation}
{\chi}_{red}^2 = {\sum_j} (O_j - M_j)^2/(\mu \delta^2)
\end{equation}
which was computed by comparing the normalized shape of the model H${\alpha}$ emission line, ${M_j}$, with the observed one,  ${O_j}$ at each of $j$ = 471 data points. 
The uncertainty, ${\delta}$, in the observed normalized line profile intensities is estimated to be 4\% based on noise in the spectrum. 
Collectively, there are three parameters of interest ( ${\xi}_i$, ${n}_{i}$, $q$ ) and
${\mu}$ degrees of freedom where ${\mu}$ =  471 data points - 3 parameters - 1. The summation was performed over the velocity span of the broad H${\alpha}$ line depicted in Figure 5.

\begin{figure}
	\includegraphics[width=\columnwidth]{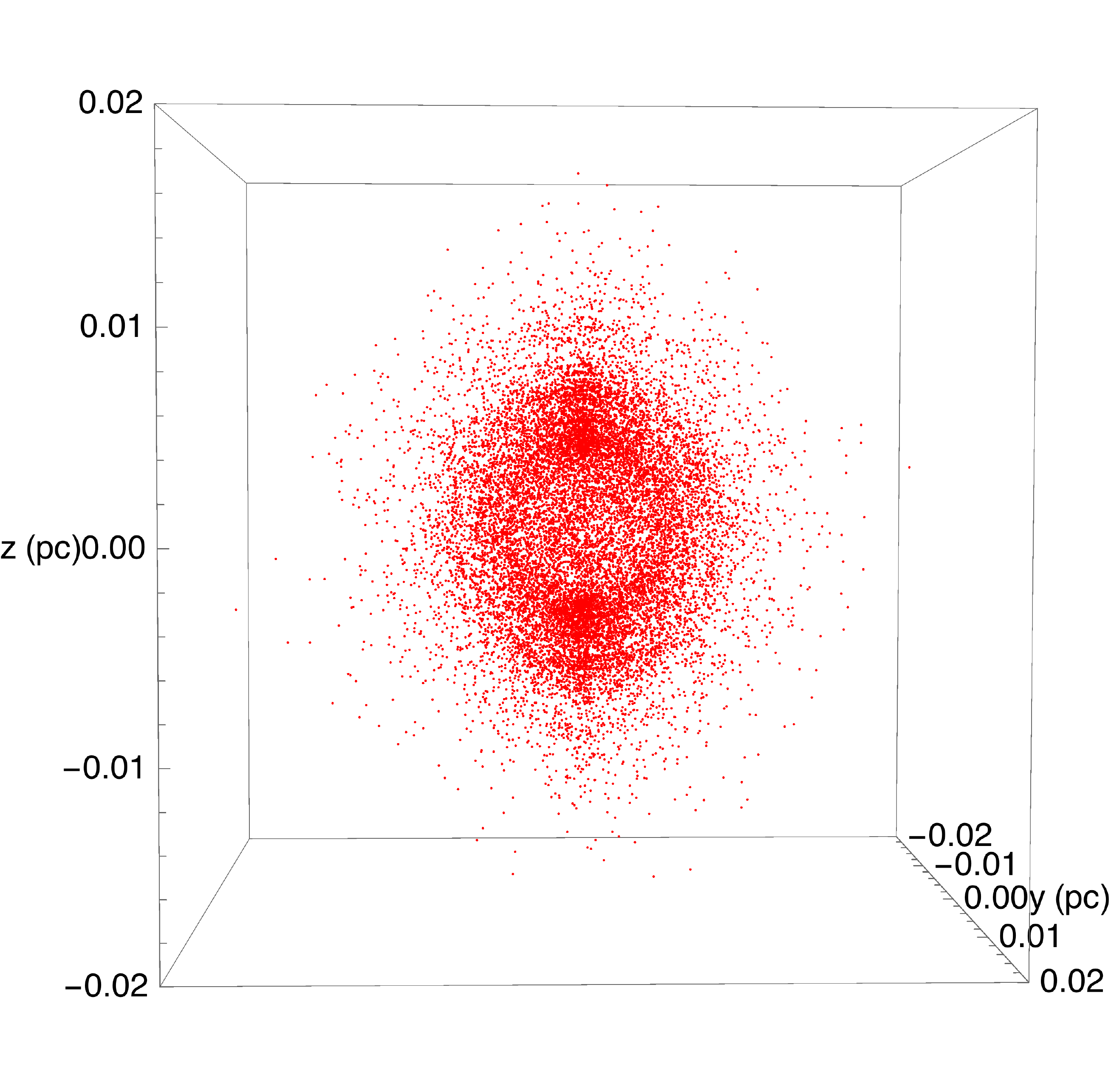}
     \caption{The spatial distribution of the 20458 points that produces the model H${\alpha}$ emission line illustrated in Fig. 5. The $y$-axis is the line-of-sight along which the radial velocities were computed and the $z$ axis is in the sky plane. The box containing the points is 0.04 pc on the side. The figure is a frame in a 3-D interactive visualization that is accessible in the online journal. } 
   \label{fig:example_figure}
\end{figure}

\subsubsection{Radial motion at the escape velocity}

The {\scriptsize XSTAR} photoionization model that best reproduces the triangular shape observed for the H${\alpha}$ emission line is defined by the {\scriptsize XSTAR} parameters log$_{10}$ ${n}_{i}$  (cm$^{-3}$) = 8.5,  log$_{10}$ ${\xi}_i$ (erg cm s$^{-1}$) = 3.3, $q$ = 0.75 corresponding to the location of minimum ${\chi}_{red}^2$ =1.5 in Fig. 4. The 
resulting model H${\alpha}$ emission line shape (Fig. 5) depicts the line-of-sight velocity for 20458 points moving radially at the escape velocity

\begin{equation}
v_{esc} = \sqrt{2GM(r)/r}
\end{equation}
where $M(r)$ is the sum of the BH mass and the surrounding stars (D13), interior to an ionization bounded spherical volume that has an outer radius ${\sim}$ 0.02 pc. Visually, the resulting distribution of points looks rather like the Greek symbol ${\phi}$ as illustrated in Fig. 6.  The Balmer emissivity determines most of the visible structure, including the radial extent of the points and the bright ring, whereas the linear feature illustrates the crowding of points at the poles, a consequence of heuristically distributing the points randomly in ${\theta}$, as opposed to  ${\sin\theta}$ in order to obtain a triangular emission line profile shape. A symmetric emission line results from a bi-symmetry in the line-of-sight velocity, depicted for a subset of the points in Fig. 7, and described by

\begin{equation}
v(r, \theta,\phi) = v_{esc} \cos\theta \sin\phi
\end{equation}

\begin{figure}
	\includegraphics[width=\columnwidth]{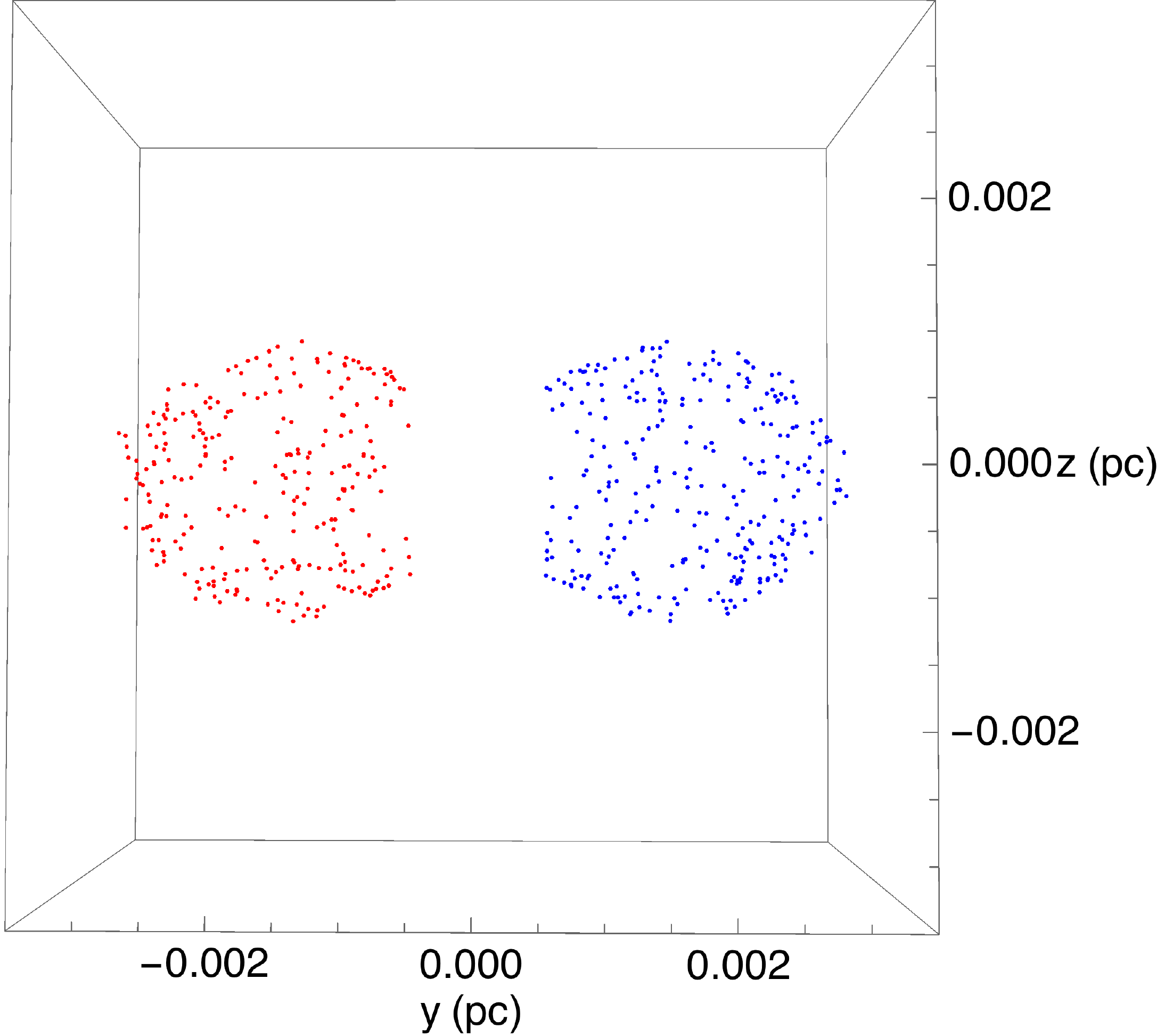}
     \caption{This view along the $x$-axis of Fig. 6 depicts an isovelocity surface that exhibits a conspicuous double-lobe shape. The illustration has been created by restricting the points plotted in Fig. 6 to a subset with radial velocities in the range 3250 ${\le}$ |$v$| (km/s) ${\le}$ 3500. The points are colour-coded red and blue according to their radial velocity measured along the line-of-sight ($y$-axis). The red lobe would be pointed at the observer for an inflow, and conversely, the blue lobe would be pointed at the observer for an outflow. The model H${\alpha}$ emission line illustrated in Fig. 5 represents the superposition of many isovelocity surfaces that span the full range of observed radial velocities. The box containing the subset of points is 7 ${\times}$ 10$^{-3}$ pc on the side. The figure is a frame in a 3-D interactive visualization that is accessible in the online journal.} 

   \label{fig:example_figure}
\end{figure}
Isovelocity contours manifest as double lobes when viewed along an axis that is perpendicular to the line-of-sight, one lobe is redshifted and the other blueshifted. However, when viewed along the line-of-sight, the two lobes appear to be superimposed with the red-lobe nearest the observer for an inflow, and vice-versa for an outflow. The model H${\alpha}$ emission line illustrated in Fig. 5 represents the superposition of many such lobes, or more precisely, nested isovelocity surfaces that span the full range of observed velocities.

\subsubsection{Dust extinction to the BLR}
 
The {\scriptsize XSTAR} photoionization model described above also predicts the emergent luminosities for numerous emission lines including H${\alpha}$, H${\beta}$, H${\gamma}$ and \ion{He}{ii} ${\lambda}$1640, which can be directly compared with {\it HST} observations to infer dust extinction to the BLR (Table 1). Including only the bright H and He lines and excluding several bright metal lines is a hedge against the unknown metallicity for the BLR gas.  Besides, the bright permitted UV emission lines, \ion{Mg}{ii} ${\lambda}$2798, \ion{C}{iv} ${\lambda}$1548 and Ly${\alpha}$ are too strongly self-absorbed to be of any use for quantitative analysis.
The results, illustrated in Fig. 2, show that at the wavelength of \ion{He}{ii} ${\lambda}$1640, the dust extinction to the BLR gas is ${\sim}$ 3.3 mag, identical to that inferred using the ADAF continuum (Section 2.2). The extinction diminishes considerably to an average ${\sim}$ 0.5 ${\pm}$ 0.2 mag in the visible 
based on the broad H${\alpha}$, H${\beta}$, H${\gamma}$  emission lines. H${\delta}$ was not included in the analysis because it is faint and potentially blended with ${[}\ion{S}{ii}{]}$ (D13).
Overall, the dust extinction to the broad emission lines in the visible is consistent with being independent of wavelength with the remaining uncertainty due to the model not reproducing the extinction corrected Balmer emission line luminosities to an accuracy better than ${\sim}$ 15\%. 

\begin{table}
\centering
\caption{Bright H and He recombination lines used to define dust extinction to the H$^{+}$ region}
\begin{threeparttable}
\begin{tabular}{cccc}
\hline
Line\tnote{\it a}  & Observed & {\scriptsize XSTAR} & Extinction   \\
 & 10$^{40}$ erg s$^{-1}$ & 10$^{40}$ erg s$^{-1}$ & mag \\
(1) & (2) &  (3)  & (4)  \\
\hline
H${\alpha}$ ${\lambda}$6564 & 9.54 & 19.74 & 0.79 \\
H${\beta}$ ${\lambda}$4862 &  2.40 & 3.70 & 0.47 \\
H${\gamma}$ ${\lambda}$4341\tnote{\it b} & 0.76 & 1.04 & 0.34 \\
${\ion{He}{ii}}$ ${\lambda}$1640  & 0.06 & 1.38 & 3.27 \\
\hline
\end{tabular}

\begin{tablenotes}\footnotesize
\item[\it a] Quoted wavelengths are rest-frame vacuum in {\AA}
\item[\it b] H${\gamma}$ is blended with ${[}\ion{O}{iii}{]}$${\lambda}$4364 the incomplete subtraction of which may bias the extinction estimate. 
\end{tablenotes}
\end{threeparttable}
\end{table}

\subsubsection{Randomly oriented circular orbits}

For the purposes of comparison, another grid of model H${\alpha}$ emission line profiles was computed covering the same (${n}_{i}$, ${\xi}_i$, $q$) parameter space as illustrated in Fig. 4 
but employing 
a different kinematic law, one that describes randomly oriented circular orbits. Comparing each model H${\alpha}$ emission line with the observed one yielded a minimum ${\chi}_{red}^2$  = 1.5 corresponding to the {\scriptsize XSTAR} model parameters log$_{10}$ ${n}_{i}$  (cm$^{-3}$) = 8.75,  log$_{10}$ ${\xi}_i$ (erg cm s$^{-1}$) = 3.05, and $q$ = 0.35. By comparison with radial motion at the escape velocity, randomly oriented circular orbits require an H$^+$ region with a combination of higher density and smaller radius to widen the model H${\alpha}$ emission lines sufficiently to compensate for the factor of ${\sqrt {2}}$ intrinsically smaller FWHM. Although the higher density model produced an H${\alpha}$ emission line shape that is statistically indistinguishable from the observed one illustrated in Fig. 5, {\scriptsize XSTAR} also predicts luminosities for the H${\alpha}$, H${\beta}$, H${\gamma}$ and \ion{He}{ii} ${\lambda}$1640 emission lines that, by comparison with the {\it HST}/STIS observations, imply a shape for the dust extinction curve that is steep in the visible and flat in the UV.
Such an unusual shape for the dust extinction curve has been inferred for other AGN \citep{Gaskell_2004}. However, the predicted reddening in the visible, corresponding to $E(B-V)$ = 0.75 mag, is inconsistent with the observed H${\alpha}$/H${\beta}$ ratio (Fig. 3, Section 2.3). Consequently, this higher density {\scriptsize XSTAR} model involving randomly oriented circular orbits is not considered further.

\section{Discussion}

Analysis of the broad H${\alpha}$ emission line in NGC 3327
(Section 2.4) suggests that the BLR has little angular momentum. This is because the kinematic model that best describes the broad H${\alpha}$ emission line shape involves radial motion at the escape velocity. The question as to whether the radial motion
represents an inflow or an outflow can be addressed using the equation of hydrostatic equilibrium, which entails understanding the physical conditions inside the H$^+$ region.

\begin{figure}
	\includegraphics[width=\columnwidth]{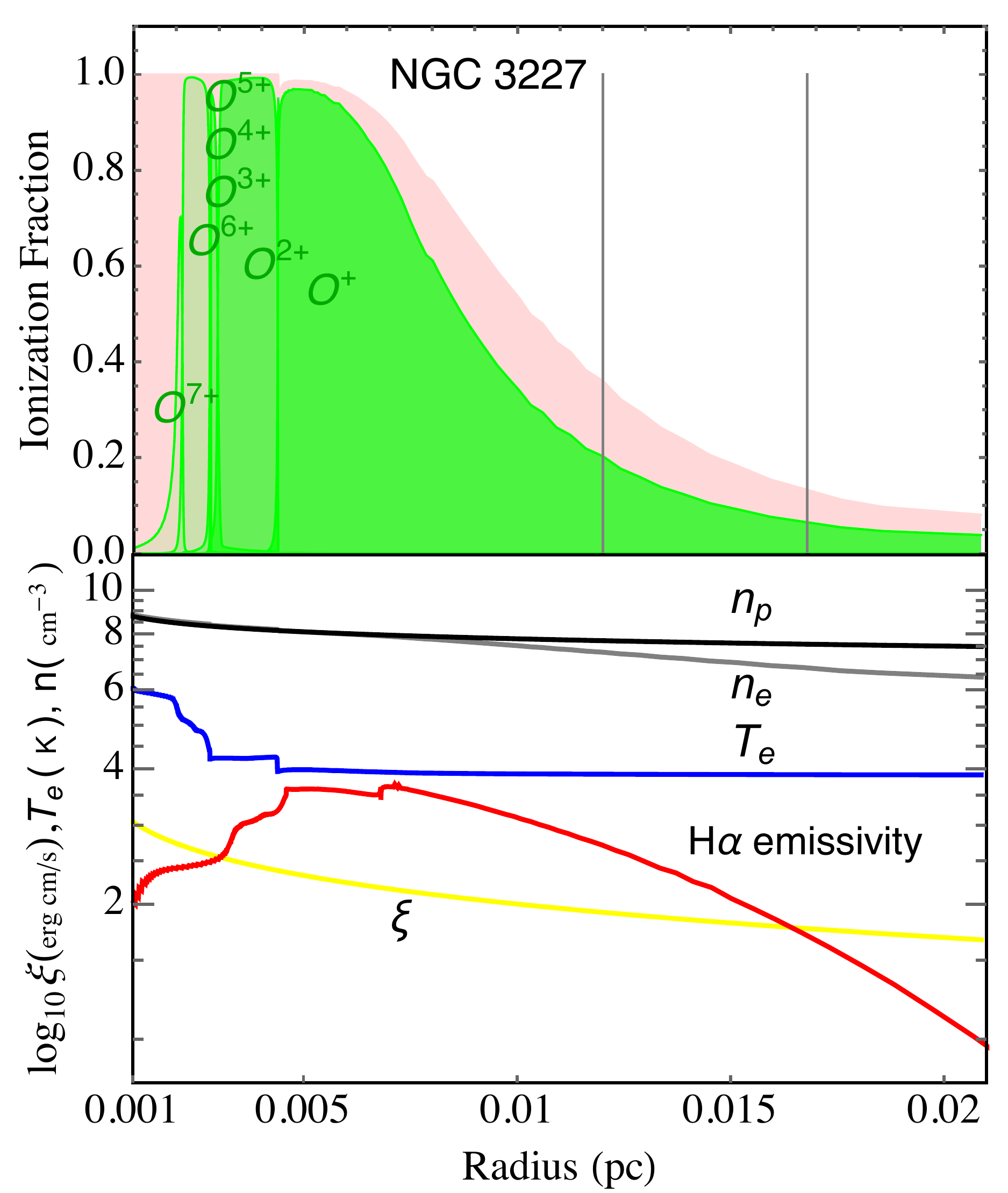}
     \caption{The upper panel illustrates the radial dependence of the ionization fraction, indicated on the ordinate, for H (pink shading), and O (green shading). 
     The two vertical lines identify two measures of the dust reverberation radius (Section 3.1).  The lower panel illustrates the radial distribution of the 
     proton $n_p$ (black line), and electron $n_e$ (gray line) number densities, the electron temperature (blue line), the arbitrarily
     scaled H${\alpha}$ emissivity (red line), and the ionization parameter, ${\xi}$ (yellow line). The ordinate identifies the relevant units, and the abscissa indicates
     radial distance from the BH in pc for both the upper and lower panels.   } 

   \label{fig:example_figure}
\end{figure}

\subsection{Physical conditions within the H$^+$ region}

The broad H${\alpha}$ emission line has a full-width at zero-intensity (FWZI) corresponding to ${\sim}$ 7000 km/s rest-frame (Fig. 5). This finite width can be understood in terms of a precipitous decline in the Balmer emissivity 
that is caused by an equally dramatic increase in the electron temperature as the central UV--X-ray source is approached. These trends are illustrated in Fig. 8 along with the radial dependence for several other important physical parameters.

The inner radius of the H$^+$ region is coincident with the empirically determined Balmer reverberation radius \citep{De_Rosa_2018}, corresponding to a physical size of 10$^{-3}$ pc. At this radius H is fully ionized with an electron temperature of 10$^{6}$ K and a number density of 3 ${\times}$ 10$^{8}$ cm$^{-3}$, but
the ionization fraction decreases dramatically with increasing radius. As a result the H$^+$ region is partially ionized over 98\% of its volume, a phenomenon caused 
by the large H column, N$_H$ = 5.5 ${\times}$10$^{24}$ atoms cm$^{-2}$, which presents a formidable opacity to H ionizing photons. 
Empirically,  the inverse correlation between radius and filling factor for ${\ion{H}{ii}}$
regions reported by \cite{Cedr_s_2013} indicates a filling factor of unity for nebulae with radii smaller than ${\sim}$ 1 pc which, with an outer radius
of 0.02 pc, would include the BLR in NGC 3227.

The dust reverberation radius has been measured to be in the range 14 to 20 light-day \citep{Suganuma_2006, Koshida_2014},
placing the onset of dust at the periphery of the partially ionized zone (Fig. 8). Evidently, the dust must be distributed in a low inclination ring 
so as to not affect the Balmer emission lines (Fig. 3, Section 2.3).
Evidence that dust is severely depleted inside the H$^+$ region, presumably by sublimation \citep{Barvainis_1987}, is provided by the modest dust extinction. Compared to the large
H column, the dust extinction, corresponding to 0.5 mag in the visible (Fig. 2), is about five orders of magnitude smaller than would be expected for a 
Galactic gas to dust mass ratio of ${\sim}$ 100 \citep{Spitzer_1978}.

{\scriptsize XSTAR} predicts O to be progressively ionized through all its possible ionization stages as the central UV--X-ray source is approached providing a natural explanation
for the absence of broad forbidden ${[}\ion{O}{iii}{]}$ emission lines.
No ${[}\ion{O}{ii}{]}$ ${\lambda}$${\lambda}$3727, 3729 emission lines appear in the {\it HST} spectra, consistent with the {\scriptsize XSTAR} model. Furthermore,
both the observed reddening-insensitive ratio ${[}\ion{O}{iii}{]}$${\lambda}$$5008/{[}\ion{O}{iii}{]}$${\lambda}$4960 and the value predicted by {\scriptsize XSTAR} are
consistent with the theoretical  value \citep{Dimitrijevic_2007}. However, there are two notable discrepancies between the {\it HST} observations and 
the {\scriptsize XSTAR} model pedictions. The first is that the ${[}\ion{O}{iii}{]}$${\lambda}$5008 emission line is observed to be 230\% brighter than
predicted by {\scriptsize XSTAR}. This excess emission could be attributed to the extended source of ${[}\ion{O}{iii}{]}$${\lambda}$5008 seen in the nucleus of NGC 3327 \citep{Arribas_1994, Mundell_1995, Delgado_1997, Garc_a_Lorenzo_2001, Walsh_2008}. However, the second discrepancy, that {\scriptsize XSTAR} predicts the ${[}\ion{O}{iii}{]}$${\lambda}$4364 emission line to be ${\sim}$ 25 times brighter than the upper limit estimated in D13, is not as easily addressed as the first. On the one hand the ${[}\ion{O}{iii}{]}$${\lambda}$5008/${\lambda}$4364 ratio produced by {\scriptsize XSTAR} is entirely consistent with that expected \citep{Osterbrock_1989} given the temperature ${\sim}$10$^4$ K and electron density ${\sim}$10$^8$ cm$^{-3}$ in the O$^{++}$ region. On the other hand, the observed ${[}\ion{O}{iii}{]}$${\lambda}$4364 emission line would have to be as bright as H${\gamma}$ in order to match the {\scriptsize XSTAR} prediction. Observationally, this seems unlikely, but the resolution of the ${\it HST}$ observations is insufficient to adequately resolve the two lines in question (D13). Equally disconcerting, however, is that contrary to commonly accepted wisdom, {\scriptsize XSTAR} predicts the ${[}\ion{O}{iii}{]}$${\lambda}$4364 emission line to be bright even though the electron density in the O$^{++}$ region exceeds the critical density of the ${^1S_0}$ level responsible for the transition by about one order of magnitude.

X-ray observations reported by \citet{Markowitz_2009} reveal the He-like $\ion{O}{vii}$ emission line triplet, but ratios of the line strengths are unable to distinguish between photoionization and collisional excitation. {\scriptsize XSTAR} predicts the forbidden line to be the brightest, as would be expected in a photoionized gas, but about a factor of 3 brighter than observed. {\scriptsize XSTAR} further predicts the intercombination line to be the same brightness as observed, and the resonance line, for which the measurement uncertainty is large, to be about a factor of 4 brighter than observed.

\subsection{Dynamics of the H$^+$ region}

The BLR dynamics can be explored by establishing whether or not the gas and radiation pressure gradients, $\frac{dP(r)}{dr}$, are sufficient to maintain hydrostatic equilibrium 
according to the relation,

\begin{equation}
\frac{dP(r)}{dr} = \frac{G M(r) \rho(r) }{r^{2}}
\end{equation}
where $G$ is the gravitational constant, $M(r)$ is the sum of the BH mass plus the mass of surrounding stars (D13), $r$ is the radial distance from the central BH, and ${\rho}(r)$ is the 
gas volume density (Fig. 8). The condition for hydrostatic equilibrium is illustrated in Fig. 9.

The gradient in the gas pressure is

\begin{equation}
\frac{dP(r)}{dr} = \frac{d [2 n(r) k T(r)]}{dr}
\end{equation}
where $T(r)$ and $n(r)$ represent the radial dependence of the electron temperature and number density illustrated in Fig. 8. The gradient was calculated by numerically differentiating the 
{\scriptsize XSTAR} gas pressure. The result, shown in Fig. 9, is not smooth because the finite resolution of the {\scriptsize XSTAR} computations causes
the gas pressure to exhibit occasional step functions which transform into delta functions when differentiated.

The gradient in the ionizing radiation pressure is

\begin{equation}
\frac{dP(r)}{dr} = \frac{ L_{ion}}{2 {\pi} r^3 c}
\end{equation}
with $c$ representing the speed of light.
\begin{figure}
	\includegraphics[width=\columnwidth]{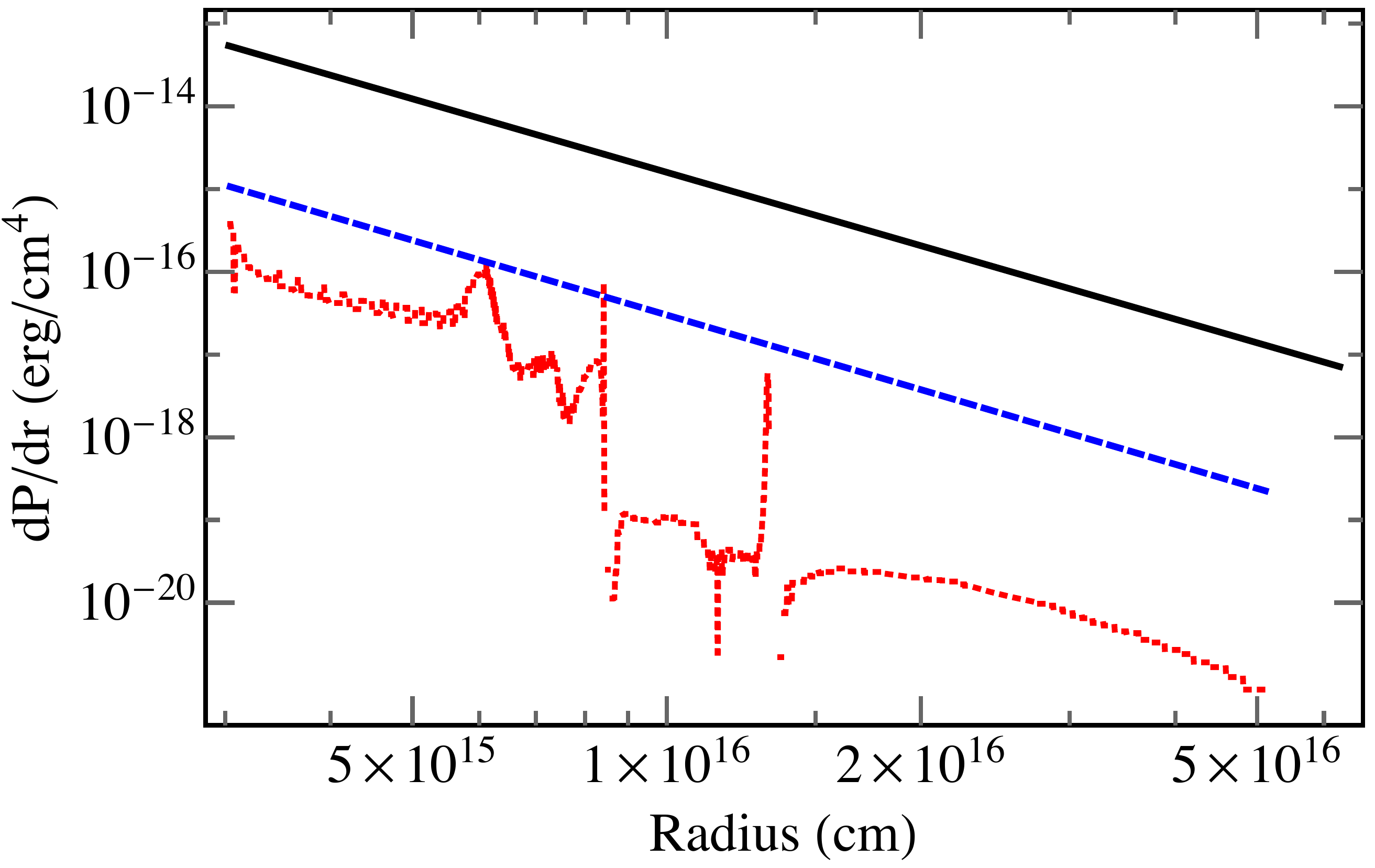}
     \caption{The gradient in the gas pressure (red-dotted line), the gradient in ionizing photon radiation pressure (blue-dashed line), and the condition for hydrostatic equilibrium (black-solid line). Radial distance from the BH is indicated on the abscissa in units of cm, pressure gradient is indicated on the ordinate in units of erg cm$^{-4}$. } 
   \label{fig:example_figure}
\end{figure} 
Fig. 9 shows that neither the gas or radiation pressure gradients, nor their sum, is sufficient to maintain hydrostatic equilibrium in such close proximity to the BH. 
This statement is likely to be true even allowing 
for the force generated by line-absorption of radiation.  The calculations of \cite{Stevens_1990} predict that the radiation force multiplier will be less than unity in X-ray 
illuminated stellar winds with 2 ${\le}$ log$_{10}$ ${\xi}$ ${\le}$ 3, ionization parameters that overlap the range inferred for the BLR of NGC 3227 (Fig. 8).
Consequently, inflow of the BLR gas at the free-fall velocity is inevitable. 

Assuming spherical symmetry, integrating the H column reveals that the BLR contains a mass $m$ = 40 M$_{\sun}$ of partially ionized H gas. 
The mass inflow rate $\dot{m}$ can be estimated by integrating the equation of continuity
\begin{equation}
\dot{m}  =  \epsilon 4{\pi} r^2 n_p(r) m_p v_{esc} 
\end{equation}
where the filling factor ${\epsilon}$ = 1, and the proton number density ${n_p}$       
= 5.62 ${\times}$10$^8$ cm$^{-3}$ at the inner radius, ${r_i}$ = 10$^{-3}$ pc, where $v_{esc}$ = 5966 km/s. The proton mass $m_p$ = 1.67 ${\times}$ 10$^{-27}$ kg leading to 
$\dot{m}$ = 1 M${_{\sun}}$/yr, superseding a prior estimate in D13.
Assuming radiatively inefficient accretion, the X-ray luminosity requires 
${\sim}$ 10$^{-2}$ M$_{\sun}$ yr$^{-1}$ (D13). Thus, if one assumes steady-state mass conservation, 99\% of the inflowing gas must be re-directed into an outflow \citep{Barbosa_2009,Alonso_Herrero_2019}, and on a very short timescale indeed corresponding to ${m/\dot{m}}$ ${\sim}$ 40 years. Alternatively, the X-ray luminosity has been underestimated which is a distinct possibility if the BLR is indeed Compton thick as implied in Section 3.1. Another potential resolution of the apparent discrepancy between the inflow and BH accretion rate is that the radiative efficiency of the ADAF has been overestimated by one or two orders of magnitude \citep{Blandford_1999}. Regardless, radial inflow 
provides a straightforward way out of the angular momentum conundrum that has stymied our understanding of AGN fueling for several decades \citep[][and references therein]{Begelman_1994,Hobbs_2011}. 

Geometrically, the BLR in NGC 3227 is bisymmetric, exhibiting two hemispheres of inflowing gas (Fig. 6) with the redshifted one, depicted in Fig. 7, facing the observer.
This combination of kinematic bisymmetry and orientation causes the triangular shape observed for the broad H${\alpha}$ emission line. The low inclination inferred for the dust torus \citep{Alonso_Herrero_2011} and the spectroscopic designation as a Seyfert 1 suggest an almost face-on aspect. In this context, the double radio source identified by \cite{Mundell_1995} may actually be the combination of a core and Doppler-boosted one-sided jet pointing roughly in our direction.

\subsection{X-ray emission versus absorption}

On the one hand, the implication that the BLR of NGC 3227 is Compton-thick (Section 3.1) is not supported by observations of high energy X-ray absorption \citep{Panessa_2006}.
On the other hand, comparing the FWHM of the Fe K${\alpha}$ emission line reported by \cite{Markowitz_2009} to the FWZI of the broad H${\alpha}$ line (Section 3.1) 
indicates that much of the Fe K${\alpha}$ emitting region lies inside the Balmer reverberation radius. Although the physical properties of that X-ray emitting region are not described specifically by the {\scriptsize XSTAR} modelling reported here, one can infer, by extrapolating the electron density inward from the Balmer reverberation radius, 
that the requisite column needed to produce the Fe K${\alpha}$ line should exist, corresponding to ${\sim}$ 10$^{23}$ atoms cm$^{-2}$.

\section{Conclusion}

The {\scriptsize XSTAR} photoionization modelling results reported here reveal the physical properties of the BLR in NGC 3227. Given the small emitting volume ${\sim}$ 10$^{-6}$ pc$^{3}$, the ${\sim}$ 10$^7$ L$_{\sun}$ broad H${\alpha}$ emission line requires ${\sim}$ 40 M$_{\sun}$ of H gas with a density ${\sim}$ 10$^8$ cm$^{-3}$ and a filling factor of unity. Additionally, the bisymmetric shell of dust-free photoionized gas is compressed, by gravity, into a Compton thick column that is dynamically unstable to infall. 
In conclusion, the Seyfert 1 nucleus of NGC 3227 appears to represent the epitome of the situation described by \cite{Pringle_2007} whereby the gravitational force
overwhelms the thermal and radiation pressure forces, allowing gas to free-fall supersonically onto the central supermassive BH. 

\section*{Acknowledgements}

I am very grateful to several of my colleagues for help along the way. In particular, Dr. Steven Willner for insight, advice, and for carefully editing the manuscript.
Prof. Ski Antonucci for kindly critiquing, with surgical precision, my ``radical ideas" expressed in said manuscript.
Last, but not least, Prof. Matt Malkan for an interesting and charitable referee report. 
STSDAS was a product of the Space Telescope Science Institute, which is operated by AURA for NASA. Based on observations made with the NASA/ESA Hubble Space Telescope, obtained from the data archive at the Space Telescope Science Institute. STScI is operated by the Association of Universities for Research in Astronomy, Inc. under NASA contract NAS 5-26555. 

\section{Data Availability Statement}

The datasets underlying this article are listed in D13 and available in the Mikulski Archive for Space Telescopes (MAST) which is the primary archive and distribution center for {\it HST} data, distributing science, calibration, and engineering data to {\it HST} users and the astronomical community at large (https://archive.stsci.edu/hst/). 




\bibliographystyle{mnras}
\bibliography{N3227} 






\bsp	
\label{lastpage}
\end{document}